\documentclass[pdftex, 12pt]{article}
\RequirePackage[T2A]{fontenc}
\RequirePackage[cp1251]{inputenc}
\RequirePackage[russian,ukrainian, english]{babel}
\usepackage{graphicx}

\title{Parameter estimation in models generated by SDE's with symmetric alpha stable noise}

\author{D. O. Ivanenko, R. V. Pogorielov}

\usepackage{amsmath,amsfonts,amsthm,marvosym}
\usepackage{eucal,mathrsfs,dsfont}
\usepackage{graphicx,alltt}
\usepackage{paralist}
\usepackage{color,multicol}
\usepackage{lscape}
\usepackage{longtable}
\usepackage{array,tabularx,tabulary,booktabs} % Дополнительная работа с таблицами
\usepackage{longtable}  % Длинные таблицы
\usepackage{multirow} % Слияние строк в таблице
\usepackage{wrapfig} % Обтекание рисунков и таблиц текстом

\newcommand{\df}{\mathrm d}

\newcommand{\E}{\mathsf E}

\newcommand{\R}{\mathbb R}
\renewcommand{\Re}{\mathbb R}
\newcommand{\prt}{\partial}

\newcommand{\dif}{{\mathrm D}}

\newcommand{\be}{\begin{equation}}
\newcommand{\ee}{\end{equation}}

\begin{document}
	\maketitle
	\frenchspacing \thispagestyle{empty}
	\renewcommand{\figurename}{Fig.}
	\renewcommand{\tablename}{Table}
	\renewcommand{\abstractname}{Abstract}
	\begin{abstract}
		The article considers vector parameter estimators in statistical models generated by Levy processes. An improved one step estimator is presented that can be used for improving any other estimator. Combined numerical methods for optimization problems are proposed. A software has been developed and a correspondent testing and comparison have been presented.
	\end{abstract}
	
	\section{Introduction}
	Given a discretely observed process that is a solution to the stochastic differential equation

	\begin{equation}\label{eq1}
	\df X_t^\theta=A_\theta(X_t^\theta)\df t +\df Z_t,
	\end{equation}
	where $A$ is a drift-function,  $Z$ is an alpha-stable process with limited jumps, $\theta$ is unknown parameter. Observation is held with a constant step $h$. 
	
	Stochastic models generated by alpha-stable processes occur in financial modeling \cite{1}-\cite{6}, physics \cite{7}, clymatology \cite{8} etc. In \cite{8} it was shown that the fast time scale noise forcing the climate contains a component with an alpha-stable distribution. Models, genereated by stochastic differential equations often contain unknown parameters that have to be estimated. In this article SDE driven by alpha-stable noise where drift function has unknown parameter is considered.
	
	It has been presented combined numerical methods for numerical estimation. These methods are used for solving systems of equations and optimization problems. They can improve convergence and accuracy in parameter estimation and do not require using high-order derivatives. 
	
	%It was shown in the article [2] that in terms of Levy's measure an algorithm for setting the limits of efficiency can be implemented. In this article, the algorithm was purely theoretical and implemented on a concrete example. In addition, there was a roughness in the transition from conditional expected value to unconditional in the algorithm. In this article, this roughening has been fixed, and the result has been generalized to a multidimensional vector parameter.
	%Section 2 presents an algorithm for checking the efficiency of the method of estimating an unknown parameter, generalized to a vector parameter. Also, the roughness correction by a bridge for conditional expected value and generalization on multiparameter models will be presented. Also, the second section will introduce a method for evaluating the parameter which effectiveness will be checked.
	
	%As this model contains a stochastic integral over the Poisson measure, ordinary numerical methods are not suitable, so sections 3 and 4 will present numerical methods that will be used to evaluate the parameter for a particular task. The simulation of increments for integrals was considered in [3]
	
	%Section 5 will present the results obtained and describe the software that was developed in the process of solving this problem.
	
	The structure of paper is following. In the beginning it is presented general information and explanation of models. An algorithm for checking the efficiency of the estimating method has been presented for scalar case. This algorithm gives a possibility to compare an ammount of statistical information loss. Next it is described estimators that have been considered with methods of their obtaining. Then it is given a description of numerical methods that have been used for various estimators and their comparison. Finally it is presented numerical results and the software description that has been developed during the process of problem solving. 
	
	\subsection*{Acknowledgements} This research was partially supported by the Alexander von Humboldt Foundation within the Research Group Linkage Programme {\it Singular diffusions: analytic and stochastic approaches} between the University of Potsdam and the Institute of Mathematics of the National Academy of Sciences of Ukraine.

	\section{General information and explanation}
	Consider process $X$ that is solution to the equation 1. Assume that the alpha-stable process with jumps $Z$ has Levy-Ito decomposition:
	
	$$
	Z_t=ct+\int_0^t\int_{|u|>1}u\nu(\df s, \df
	u)+\int_0^t\int_{|u|\leq 1}u\tilde\nu(\df s, \df u),
	$$
	where $\nu$ is the Poisson point measure with compensator $\df s \mu(\df u)$ , $\tilde \nu (\df s, \df u)=\nu(\df s, \df
	u)-\df s \mu(\df u)$ is corresponding compensated measure. Following investigations in \cite{9}-\cite{13} it is assumed that $\nu$ satisfies the following conditions:
	\begin{itemize}  \item[(i)] For some $\kappa>0$,
		$$
		\int_{|u|\geq 1}u^{2+\kappa}\mu(du)<\infty;
		$$
		
		\item[(ii)] For some $u_0>0$, the restriction $\mu$ on $[-u_0,
		u_0]$ has a positive density $$\sigma\in
		C^2\left(\left[-u_0,0\right)\cup \left(0, u_0\right]\right);$$
		
		\item[(iii)] There exists $C_0$ such that 
		$$
		|\sigma'(u)|\leq C_0|u|^{-1}\sigma(u),
		$$
		$$		
		|\sigma''(u)|\leq C_0u^{-2}\sigma(u),
		$$
		$$
		\ |u|\in (0, u_0];
		$$		
		\item[(iv)] $$ \left(\log {1\over
			\epsilon}\right)^{-1}\mu\Big(\{u:|u|\geq \epsilon\}\Big)\to \infty,\quad
		\epsilon\to 0.
		$$
	\end{itemize}
	
	Besides, it is assumed in simulations and obtaining numerical results there is technical restriction that the jump value is bounded by some constant.

	The developed application implements various methods for estimating unknown parameters of the drift function of the model described above. In addition, the algorithm for evaluating the efficiency of the estimation method for a scalar parameter proposed in \cite{10} article has been elaborated and implemented. In this article, using this algorithm, the efficiency of each of the  methods in the scalar case is estimated.  This algorithm is based on Hajek minimax theorem \cite{14}. The detailed scheme of the algorithm is as follows. If $\hat{\theta}$ is estimator of $\theta$ then it is possible to calculate $\sqrt{J(\theta)/I(\theta)}$ that is relative efficiency to theoretical Hajek bound. Here $I(\theta)$ is bound of normed Fisher matrices, and $J(\theta)$ is sample mean. The algorithm in case of one-parameter model is following:

	\begin{itemize}  \item[(1)] To select the method of evaluation and build an estimation $\hat{\theta}_{n}$ unknown parameter $\theta_{0}$
		\item[(2)] To generate $N$ trajectories of process $X$ given equation with $\hat{\theta}_{n} = \theta_{0} $ and for each of them to build a sample size $n$
		\item[(3)] To calculate estimators $\hat{\theta}_{n}^{k}, k = 1, ... , N$ and to find a sample variance 
		$s_{N}^2 = \frac{1}{N} \sum_{k = 1}^{N} (\sqrt{n}(\hat{\theta}_{n}^{k}- \theta)^2) $
		\item[(4)] To calculate $n_{0}$ and to generate $N$ trajectories of process $X$ з $\hat{\theta}_{n} = \theta_{0} $, for each of them to calculate $\varXi_{n}^{k}(n_{0}), k = 1,...,N$
		\item[(5)] To find a sample mean $J_{n}(\hat{\theta}_{n}, n_{0}) 
		= \frac{1}{N} \sum_{k =1}^{N}(\varXi_{n}^{k}(n_{0}))^2$
		\item[(6)] By value $\sqrt{J_{n}(\hat{\theta}_{n}, n_{0})s_{N}^2 }$ to make a conclusion about the efficiency of the method 
	\end{itemize}
	
	There are weaknesses in this algorithm. In the second step of the algorithm the parameter is estimated $N$ times. 
	In the fifth step of the algorithm there was a significant roughness in the transition from conditional expectation to unconditional one. This disadvantage is eliminated by the bridge: if the trajectory falls into the neighborhood point, then it was taken for analysis. In this article, technical and algorithmic problems have been solved for the practical implementation of this algorithm, as well as this algorithm is generalized to the case of a vector parameter.
	
	To implement the algorithm, we use the integral representation of the transition probability density obtained using the Malliavin calculus in the article \cite{13}. Below there are items and their integral representations that figure in Fisher's information matrix estimation. By the Theorem \cite{12} under the assumptions about $Z$ at point $t$ is twice differentiable and correspondent stochastic derivatives given by the formulas:
	$$DZ_t =\int_0^T\int_{\Re}\varrho(u)\nu(ds,du)$$
	$$DDZ_t =\int_0^T\int_{\Re}\varrho(u)\varrho'(u)\nu(ds,du)$$
	The Scorohod integral 
	$$\delta(1)=-\int_0^T\int_{\Re}\dfrac{(\sigma(u)\varrho(u)'}{\sigma(u)}\tilde\nu(\df s, \df u)$$ 
		
	For differentiability with respect to a parameter and the existence of stochastic derivatives process $X$ we need to involve additional restrictions about A. 
	In the one parameter case, drift has to satisfy the following conditions:
	\begin{itemize}  \item[(v)] Let a have bounded derivatives $\prt^{i+j}_{x^{i}\theta^{j}} A$, $i\leq3,j\leq2$ 	
		
		\item[(vi)] Derivatives $\prt_{x}A$, $\prt^2_{xx}A$,$\prt^2_{x\theta}A$,
		$\prt^3_{xxx}A$, $\prt^3_{xx\theta}A$, $\prt^3_{x\theta\theta}A$, $\prt^4_{xxx\theta}A$ are bounded and
		$$|A_\theta(x)|+|\partial_{\theta}
		A_\theta(x)|+|\partial^2_{\theta\theta} A_\theta(x)|\leq
		C(1+|x|). $$
		for all  $\ \theta\in \Theta, \ x\in \R$

	\end{itemize}
	By \cite{12} under the conditions (i)--(vi) $X$ is twice differentiable wrt $\theta$. 
	and the following statement are held:
	
	 Let $t_0 > 0$ is fixed. Consider the equation \eqref{eq1} with initial condition $X_{t_0}= x_{0}$. Denote further $Y_{t}^1=\prt_{\theta}X_{t}, 
	Y_{t}^2=DX_{t},Y_{t}^3=D\prt_{\theta}X_{t},Y_{t}^4=D^2 X_{t}$. Then $\overline{Y}_{t} := (Y_{t}^1,Y_{t}^2,Y_{t}^3,Y_{t}^4)$
	is the solution of the simultaneous equations:
	\begin{equation*}
	\begin{cases}
	dY_{t}^1 = \prt_{x}A_{\theta}(X_{t})Y_{t}^1 dt + \prt_{\theta}A_{\theta}(X_{t})dt
	\\
	dY_{t}^2 = \prt_{x}A_{\theta}(X_{t})Y_{t}^2 dt+ dDZ_{t}
	\\
	dY_{t}^3 = \prt_{x}A_{\theta}(X_{t})Y_{t}^3 dt+ (\prt_{x\theta}A_{\theta}(X_{t})Y_{t}^2
	+\prt_{xx}A_{\theta}(X_{t})Y_{t}^1 Y_{t}^2)dt
	\\
	dY_{t}^4 = \prt_{x}A_{\theta}(X_{t})Y_{t}^4 dt+ \prt_{xx}A_{\theta}(X_{t})((Y_{t}^2)^2) dt + dD^2 Z_{t}
	\\
	Y^{i}_{t_{0}} = 0, i = 1, ... ,4
	\end{cases}
	\end{equation*}
	Euler's method is used for solution. Coordinates of $\overline{Y}_{t}$ are components of the formula that is needed for fourth step of algorithm implementation. The theoretical formula for correspondent functional is given by:
	\begin{equation}\label{equ2}
	  \Xi_t^1={(\prt_\theta X_t^\theta) \delta(1)\over
		\dif X_t^\theta}+{(\prt_\theta X_t^\theta) \dif^2 X_t^\theta \over
		(\dif X_t^\theta)^2}-{\dif(\prt_\theta X_t^\theta)\over \dif
		X_t^\theta}. 
	\end{equation}
	
	In the next section it will be explained some types of estimators which efficiency will be checked by the algorithm above.

	%In case of linear two-parametric model drift has the form:
	%$$A(X)= aX^2+bX $$
	%where $(a,b)$ - estimated vector parameter
	
	%For three-parametric linear model drift has the form:
	%$$A(X)= aX^3+bX^2+cX $$
	%where $(a,b,c)$ - estimated vector parameter

	\section{Parameter estimation}
	
	In this paper it is focused on the following types of estimators.

	\subsection*{$L_p$ estimators}
	In this case we consider $L_p$ estimators. For example, the functional that needs to be minimized for degree two has the form:
	$$ \theta_{n} = argmin_{\theta\in \Theta} \sum_{k=1}^{n} (X_{hk}-X_{h(k-1)} - A_\theta(X_{h(k-1)}h))^2, $$
	where $\theta_{n}$ is an estimated vector parameter, $A_\theta(X_{h(k-1)})$ is the drift function, $h$ is the distance between neighbor observations.

	For least absolute value the minimized functional has the form: 
	$$ \theta_{n} = argmin_{\theta\in \Theta} \sum_{k=1}^{n} |X_{hk}-X_{h(k-1)} - A_\theta(X_{h(k-1)}h)| $$
	
	Figures below show the examples of loss functions  with $p = 1,2$ respectively and
	
	$$A_{\theta}(x) = -ax + \dfrac{b}{\log(1+x^{2})}, T = 1000, \alpha = 1.75, h = 1$$
	(True values of parameters are $a = 1, b = 1.$)

	\begin{figure}[!h]
		\begin{minipage}[h]{0.49\textwidth}
		\caption{Loss function for L1 estimator} \label{fig1}
		\includegraphics[width=\textwidth]{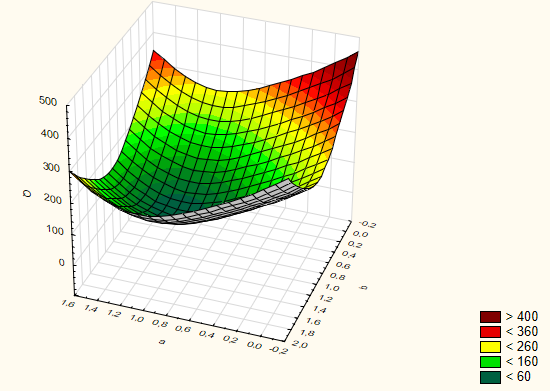}
	\end{minipage}
	\hfill
	\begin{minipage}[h]{0.49\textwidth}
		\caption{Loss function for L2 estimator} \label{fig2}
		\includegraphics[width=\textwidth]{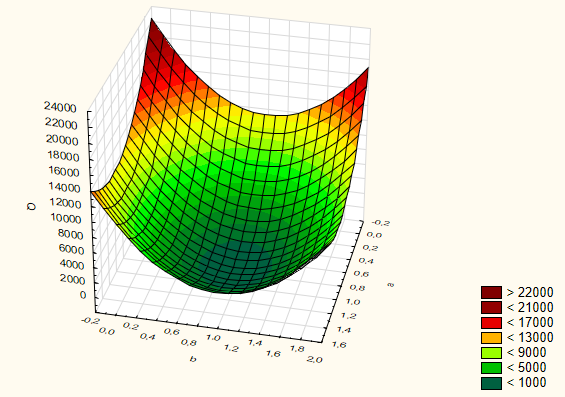}
	\end{minipage}
	\end{figure}
	
	It also considers the estimate obtained for $L^{\infty}$.
	In this case the minimized functional has the form:
	
	$$ \theta_{n} = argmin_{\theta\in \Theta} \max |X_{hk}-X_{h(k-1)} - A_\theta(X_{h(k-1)}h)| $$

	\subsection*{One step estimator}
	
	All estimators can be used in order to obtain  "one step estimator"\ that (under additional conditions \cite{15}) asymptotically tends to MLE. It will be precised the estimators above and compare the results. The general scheme for one-step estimation is:
	$$
	\theta_{one step} = \theta_{start} + H^{-1}(\theta_{start})*\nabla\ln QL_{n}(\theta_{start}),
	$$ 
	where $\theta_{start}$ can be chosen arbitrarily, for example it can be substituted by $L_p$ estimator, $H$ - Hesse's matrix for log likelihood function, $\nabla$ - gradient vector. Particulary, in the case of one dimension parameter case  estimator has the form:
	
	$$
	\theta_{one step} = \theta_{start} + \dfrac{\Xi_t^1(\theta_{start})}{\Xi_t^2(\theta_{start})}.
	$$
	Here functional $\Xi_t^1$ is given by \eqref{equ2}, and $\Xi_t^2$ is defined below by the formula \eqref{Xi_2} and is interpreted as Fisher information.
	
	In \cite{10} there was obtained  expression for second derivative of the likelihood and proven, that the logarithm of  the transition
	probability density has a second continuous derivative w.r.t.
	$\theta$ on the open subset of $(0,\infty)\times \Re\times
	\Re\times \Theta$ defined by inequality $ p^\theta_t(x,y)>0$ and,
	on this subset, admits the integral representation \be\label{log}
	\prt^2_{\theta\theta} \log
	p_t^\theta(x,y)=\E^{t,\theta}_{x,y}\Xi_t^2-\left(E^{t,\theta}_{x,y}\Xi_t^1\right)^2,
	\ee where

	\begin{multline}\label{Xi_2} \Xi_t^2:=\delta\left(\frac{1}{\dif
		X_t^\theta}\left(\delta\left(\frac{(\prt_\theta
		X_t^\theta)^2}{\dif X_t^\theta}\right)+\partial^2_{\theta\theta}
	X_t^\theta\right)\right)=\\-{1 \over \dif
		X_t^\theta}\dif\delta\left({(\prt_\theta X_t^\theta)^2 \over \dif
		X_t^\theta}\right)+{\dif\prt^2_{\theta\theta}X_t^\theta\over\dif
		X_t^\theta}+
	\\ \left({\delta(1)\over \dif X_t^\theta}+{\dif^2
		X_t^\theta \over (\dif
		X_t^\theta)^2}\right)\left(\delta\left(\frac{(\prt_\theta
		X_t^\theta)^2}{\dif X_t^\theta}\right)+\partial^2_{\theta\theta}
	X_t^\theta\right),
	\end{multline}
	with Skorokhod integral
	\be
	\delta\left(\frac{(\prt_\theta X_t^\theta)^2}{\dif
		X_t^\theta}\right)=\frac{(\prt_\theta X_t^\theta)^2 \delta(1)}{
		\dif X_t^\theta}+\frac{(\prt_\theta X_t^\theta)^2 \dif^2
		X_t^\theta}{(\dif X_t^\theta)^2} -\frac{2(\prt_\theta
		X_t^\theta)\dif(\prt_\theta
		X_t^\theta)}{\dif X_t^\theta},
	\ee
	\begin{multline*}\delta\left(\frac{(\prt_\theta X_t^\theta)^2}{\dif
		X_t^\theta}\right)= {2\prt_\theta X_t^\theta\over\dif
		X_t^\theta}\left(\delta(1)\dif(\prt_\theta
	X_t^\theta)-\dif^2(\prt_\theta X_t^\theta)\right)+{(\prt_\theta
		X_t^\theta)^2\dif\delta(1)\over\dif
		X_t^\theta}\\-{2(\dif(\prt_\theta X_t^\theta))^2\over\dif
		X_t^\theta}+ \left({\prt_\theta X_t^\theta\over\dif
		X_t^\theta}\right)^2 \left(\dif^3 X_t^\theta-\delta(1)\dif^2
	X_t^\theta\right)\\+{4\prt_\theta X_t^\theta\dif(\prt_\theta
		X_t^\theta)\dif^2 X_t^\theta\over(\dif
		X_t^\theta)^2}-{2(\prt_\theta X_t^\theta\dif^2
		X_t^\theta)^2\over(\dif X_t^\theta)^3}.
	\end{multline*}
	
	However, the calculation of second derivative is difficult and takes a lot of machine resource, so it is possible to replace it with approximations using numerical methods. One of the options for making calculations easier is Rao estimator which uses an approximation of second derivative by product of first derivatives:
	
	$$
	\theta_{one step} = \theta_{start} + \dfrac{\Xi_t^1(\theta_{start})}{\Xi_t^1(\theta_{start})^{2}}
	$$
	
The estimates obtained by the classical  and Rao methods are compared in {\bf Section 5}. Next section will present numerical methods that are used in software implementation.

	\section{Numerical aspects}

	For one-parameter estimation problem Powell's and Steffenson method have been used. The first method allows to find the minimum of the corresponding functional, and the second can be used in order to solve a system of nonlinear equations. Powell's method uses parabolic interpolation and has faster convergence than other one-dimensional optimization methods \cite{16}. Steffenson method is modification of Newton's one. According to the difference formula:
	
	$$Q'(\theta_{k})=\dfrac{Q(\theta_{k}+\delta)-Q(\theta_{k})}{h}$$
	and taking into consideration that $\delta = Q(\theta_{k})$
	it can be obtained Steffenson formula for iteration scheme:
	
	$$\theta_{k+1} = \theta_{k} - \dfrac{(Q(\theta_{k}))^{2}}{(Q(\theta_{k}+Q(\theta_{k}))-Q(\theta_{k}))}$$
	
	If drift function is polynomial e.g. $A = \sum_{i =1}^{m}(\theta_{i}x^{i})$ then LSE estimator can be obtained by solving system of linear equations (see formulae below). Direct methods of solving systems of linear equations give great rounding error, that is accumulated and matrix can be ill-conditioned. For iterative scheme a symmetric succesive over relaxation method is used (\textbf{SSOR}). The idea of the method is described in \cite{17} and here it will be shown on the example of two-parametric model. 
	
	Let $X_k$ to be solution to stochastic differential equation by Euler-Maruama scheme. Denote ${S_{ij}} = \sum_{k=0}^{T-1}X_{k}^{i}X_{k+1}^{j}$, where $T$ is the number of observations. Then system of linear algebraic equations that has to be solved in case of two-parameter system is:
	
	\begin{equation*}
	\begin{cases}
	S_{40}ha + S_{30}hb = {S_{21}} - S_{30}
	\\
	S_{30}ha + S_{20}hb = {S_{11}} - S_{20} 
	\end{cases}
	\end{equation*}

	Lets introduce relaxation parameter $\omega$. Since the matrix of coefficients with parameters is symmetric, it is possible to use the symmetric method of successive over-relaxation.  Then for this case, the iterative scheme for a system with two parameters is:
	
	\begin{itemize}  \item[(I)] Forward step: 
		\begin{equation*}
		\renewcommand{\arraystretch}{2.}
		\left\{\begin{array}{ll}
		a^{k+1/2} = (1-\omega)a^{k} + \dfrac{\omega}{S_{4}h}(\tilde{S_{2}} - S_{3} - S_{3}hb^{k} ) 
		\\
		b^{k+1/2} = (1-\omega)b^{k} + \dfrac{\omega}{S_{2}h}(\tilde{S_{1}} - S_{2} - S_{3}ha^{k} ) 
		\end{array}\right.
		\end{equation*} 
		\item[(II)] Backward step
		\begin{equation*}
		\renewcommand{\arraystretch}{2.}
		\left\{\begin{array}{ll}
		b^{k+1} = (1-\omega)b^{k+1/2} + \dfrac{\omega}{S_{2}h}(\tilde{S_{1}} - S_{2} - S_{3}ha^{k+1/2} )
		\\
		a^{k+1} = (1-\omega)a^{k+1/2} + \dfrac{\omega}{S_{4}h}(\tilde{S_{2}} - S_{3} - S_{3}hb^{k+1} ) 
		\end{array}\right.
		\end{equation*}
	\end{itemize}
	
	Using three-layer Chebyshev acceleration, finding a solution can be found by less number of iterations \cite{17}. 	
	Define system of equations $Rx = f$ that can be converted to the form $x = Gx+h$ where $x = \theta = (a,b) $ is the vector of coefficients. This can be done with the help of a certain iterative process (for example, SSOR) If the absolute value of spectral radius $\rho$ of matrix $G$ is less than 1 the given iteration process converges. So the sequence of vectors will converge to an exact solution, that is:
	
	$$\lim_{s\rightarrow\infty} \theta^{s} \rightarrow \theta^{*}$$
	
	Suppose that the $m$ iterations of the method are made and vectors  $\theta^1,\theta^2…\theta^m$ are obtained each of which is an approximation to $\theta^*$. Set the task to find the corresponding linear combination of vectors:
	
	$$\sum_{i=0}^{m} \alpha_{i}\theta^{i} = y^{m}$$
	that approaches to $\theta^*$ faster than $\theta^m.$
	As $\theta^m$ is an approximation to $\theta^*$, so $\theta^m$ has to coincide with $\theta^*$ in the case of equality of vectors on each iteration. Accordingly, the sum of the coefficients should be equal to 1.
	 Error of calculation $\theta^m$ can be written:
	$$y^m - \theta^* = p_{m}(S)(\theta^{0}-\theta^*) $$
	Here $p_{m} (G) = \alpha_{i} G^i$ is the polynomial of degree $m$, such that $p_m (1)=1$. So error depends on the spectral radius of $G$ as the smaller  radius, the less error will be. 
	
	The problem of finding this polynomial is complicated. By the Hamilton-Kelli theorem \cite{17}, it is a characteristic polynomial of $G$, for which one needs to know all eigenvalues. That's why, the problem is reduced to the polynomial search such that the spectral radius approaches to zero. 
	
	Suppose that $G$ has the following properties:
	
	\begin{itemize}  \item Its all eigenvalues are real
		\item They are in interval $[-\rho,\rho]$
	\end{itemize}
	
	Then it is possible to find a $p_m$, that is 
	
	\begin{itemize}  \item $p_m (1)=1$
		\item $max_{-\rho<\theta<\rho} |p_m (\theta)|$ has the least possible value among all polynomials of degree $m$
		
	\end{itemize}
	
	The solution to this problem uses Chebyshev`s polynomials, which are determined by the recurrence scheme:
	$$T_{0}=1,T_{1} (\theta)=\theta,T_{m} (\theta)=2\theta T_{m-1} (x)- T_{m-2} (\theta).$$
	
	The Chebyshev polynom with degree $m$ has the least deviation from zero on the interval [-1,1] among all polynomials of the same degree. 
	The three-layer acceleration of Chebyshev allows to use only three vectors : $y^{m},y^{m-1},y^{m-2}$
	Entering the coefficient $\mu_{m}=1/T_{m}(\theta/\rho)$, then $p_m (S) = \mu_{m}T_{m}(S/\rho).$
	Putting in the expression for a residual, we obtain the scheme of Chebyshev.
	Corresponding algorithm is in the following:
	\begin{itemize}  \item To determine an iterative process that will be accelerated (SSOR)
		\item  Set $\mu_{0}=1,\mu_{1}=\rho,y^{0}=\theta^{0},y^{1}=S\theta^{0} +c$
		\item To continue calculate until the required precision is obtained:
		
		$$\mu_m = \left({\dfrac{2}{\rho\mu_{m-1}} - \dfrac{1}{\mu_{m-1}}}\right)^{-1}$$
		
		$$y^{m}=\dfrac{2\mu_m}{\rho\mu_{m-1} } (Sy^{m-1}  +c)-  \dfrac{\mu_m}{\mu_{m-2}}  y^{m-2}$$	
		
	\end{itemize}
	Using Chebyshev's acceleration algorithm for the SSOR method, it is possible to reduce the number of iterations to 3 times.
	
	The problem of finding $\rho$ is solved here by the aim of SP-algorithm that is used for finding eigenvalues in symmetric matrices \cite{18}.

	Table \ref{tab1} shows a comparison of relaxation methods to evaluate the parameter for a two-parameter system with step 1, the number of trajectories 700, the value of the process parameter 1.75, start points are $(-0.3,-0.2)$

	\begin{table}[!htb]
		\begin{tabulary}{\textwidth}{|c|c|c|c|}
			\hline 
			Parameters val. & SOR & SSOR & SSOR with Chebyshev acc.\\ 
			\hline 
			-1 & -1.000000483658 & -1.000000584635 & -1.000000383673 \\ 
			\hline 
			-1 & -1.000000398645 & -1.000000048362 & -1.000000054735 \\ 
			\hline 
			Number of iter. & 8 & 6 & 5 \\
			\hline
			
		\end{tabulary} 
		\caption{Comparison of iterative methods for linear systems.}
		\label{tab1}
	\end{table}

	In general case where drift is not polynomial estimating unknown parameters is more difficult. Optimisation methods or methods of solving systems of nonlinear equations have to be used. Ordinary methods (for example fixed-point iteration, Seidel, Newton, successive over-relaxation) can't be used because it is difficult to check their convergence conditions.
	In general case a combination of Box-Wilson and Hook-Jeeves methods are used \cite{19}. 
	
	Recall that it is considered on the example of a two-parameter model. Box-Wilson method is modification of gradient method but the gradient is substituted with linear regression. The algorithm starts with factor analysis. Let $Q(a,b)$ is a minimized function and start point $\theta_{0} = (a_{0}, b_{0})$. The method begins with choosing a starting point and making trial steps to the sides. On the basis of trial steps, regression coefficients are calculated and movement towards the minimum begins. The movement continues until the value of the objective function decreases. Further at the point where the movement stops, the coefficients are calculated again and the algorithm is repeated. Table 2 shows the changes in parameter values $a,b$ and calculation on this basis. 
	\begin{table}[!htb]

		\begin{center}
			
			\begin{tabulary}{\textwidth}{|l|c|c|c|}
				\hline 
				№ & $a$ & $b$ & $Q$\\ 
				\hline 
				1 & $a_{0}-\delta$ & $b_{0}-\delta$ & $Q_{1} = Q(a_{0}-\delta, b_{0}-\delta)$ \\ 
				\hline 
				2 & $a_{0}+\delta$ & $b_{0}-\delta$ & $Q_{2} = Q(a_{0}+\delta,b_{0}-\delta)$ \\ 
				\hline 
				3 & $a_{0}-\delta$ & $b_{0}+\delta$ & $Q_{3} = Q(a_{0}-\delta, b_{0}+\delta)$ \\
				\hline
				4 & $a_{0}+\delta$ & $b_{0}+\delta$ & $Q_{4} = Q(a_{0}+\delta, b_{0}+\delta)$ \\
				\hline
			\end{tabulary} 
			\caption{Changes in parameter values.}\label{mytab2}
			
		\end{center}
	\end{table}
	
	Values of objective function are used to determine regression coefficients via formulas 
	
	$b_{1} = \dfrac{-Q_{1}+Q_{2}-Q_{3}+Q_{4}}{4}$, $b_{2} = \dfrac{-Q_{1}-Q_{2}+Q_{3}+Q_{4}}{4}$. 
	
	Regression coefficients $(b_{1},b_{2})$ are treated as the approximation of gradient and are used in iteration sheme that looks like:
	$$\theta_{j} = \theta_{0} - kqb_{j}\delta_{j},$$ where $q$ is proportion coefficient, $k$ is number of iteration, $\delta_{j}$ is step used in factor analysis, $j$ is vector parameter component. Iterations are held until the function value begins to increase. In this case, the point at which the value was minimal is taken as the starting point. After that, the algorithm steps are started again from factor analysis.
	Iterations are held until $\sqrt{b^{2}_{1} + b^{2}_{2}}>\epsilon$.	
	Method is zero-order and has fast convergence although is not very accurate. Thats why the precision needs to be improved and it is done by modified Hook-Jeeves method that is described in \cite{20}. 
	
	Table below shows comparison for least square estimation with drift function $-ax + \arctan(x^{2}+b)$, true values $(a,b)$ = (1,1), precision $\epsilon = 0.00001$ and start point $(a_{0},b_{0}) = (0.3, 1.2)$
	
	\begin{table}[!htb]

		\begin{center}
			
			\begin{tabulary}{\textwidth}{|c|c|c|c|c|c|}
				\hline 
				Method & Brown & Broyden & Box-Wilson & Hook-Jeeves & Hybrid\\ 
				\hline 
				a & 1.28737 & 1.02898 & 0.95117 & 1.00665 & 1.00895 \\ 
				\hline 
				b & 0.93876 & 1.22267 & 0.72833 & 1.05415 & 1.00541 \\ 
				\hline 
				number of iterations & 434 & 87 & 25 & 71 & 15 \\			
				\hline
			\end{tabulary} 
			\caption{Comparison of numerical methods for parameter estimating}\label{tab:mytab3}
			
		\end{center}
	\end{table}
	
	The $L^{\infty}$ estimator is special case of discrete minimax problem. A well-known fact is that the problem
	
	$$\min \max Q_{i}(\theta),\qquad Q_{i}=|X_{i}-X_{i-1} - A_\theta(X_{i-1})| $$
	can be transformed in nonlinear programming problem:
	
	$$ \min z\quad s.t. |Q_{i}|<z. i = 0,...,T
	$$
	Most of the methods use sequential quadratic programming (SQP), penalty or barrier function \cite{21}-\cite{26}. These methods have fast convergence, however, there is a difficulty in transforming minimax problem. Because of such transformation a $2(T-1)$ constraints would be obtained and using SQP will lead to solution of high-dimensional system of equations. That's why an Armijo algorithm is proposed that uses linear search \cite{27} and does not require high order objective function derivatives.

	In order to reduce the ammount of constraints the Euclidian approximation can be used [28]
	$$|\tilde{x}|=\sqrt{x^{2}+\beta^{2}}$$  ($\beta$ is small enough).

	\section{Results}
	
	A software has been developed to test the methods of estimating unknown parameters and the efficiency of estimators. The user can choose drift function $A$, number of observation $T$, time step $h$, parameters for alpha-stable process, and different parameters that are used in efficiency checking algorithm. Drift function is written by symbolic line.
	
\begin{figure}[!htb]
		\caption{Estimation for three-parameter model.}
		\includegraphics[width=\textwidth]{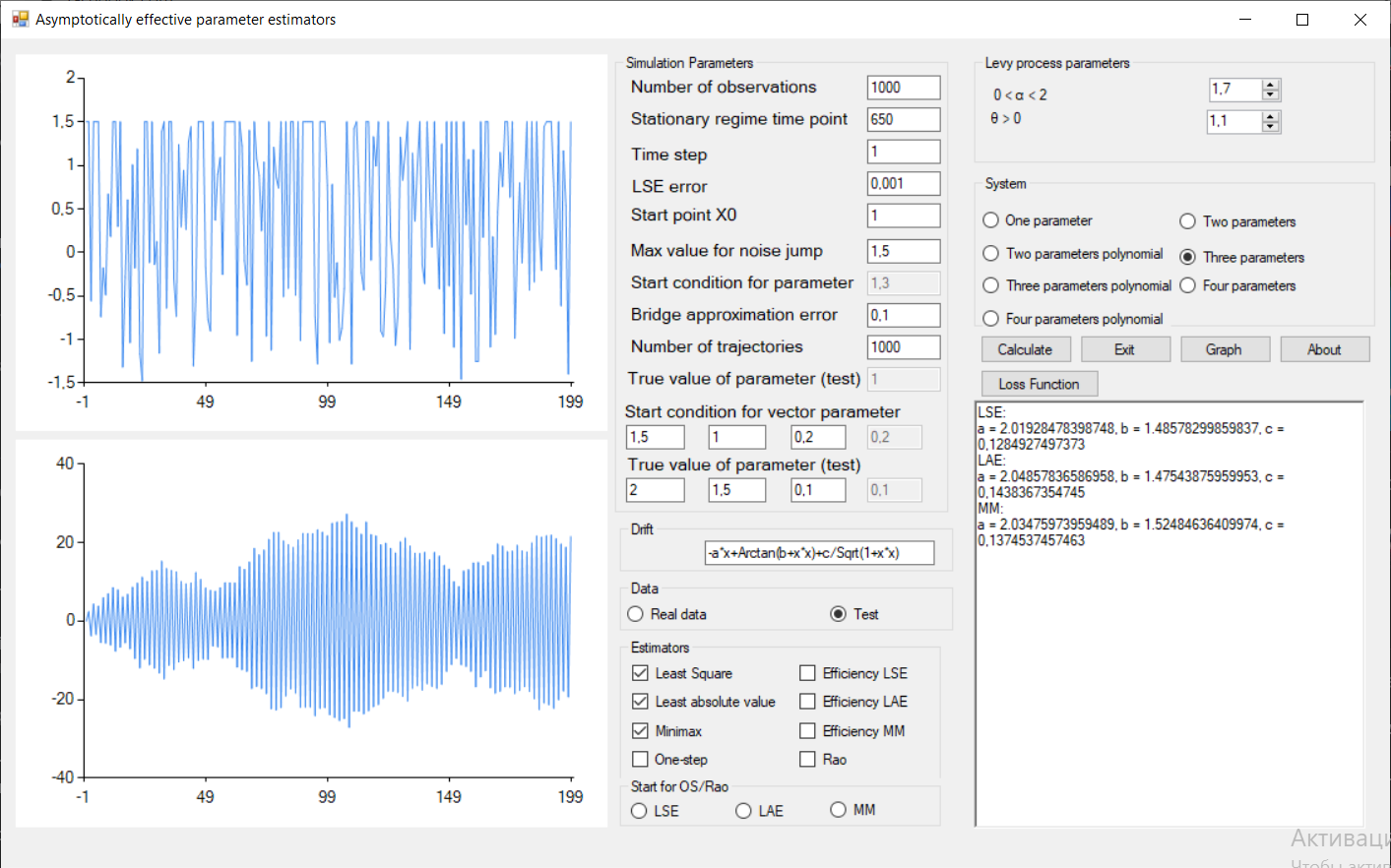}
		\label{fig3}
	\end{figure}

	Below are examples of how the application works with: $$h = 1, \alpha = 1.75, T = 1000, X_{0} = 1.$$	
	Figure \ref{fig3} shows estimation for three-parameter model with 
	$$A(x,\theta)=-a x+\arctan(b+x^{2})+\dfrac{c}{\sqrt{1+x^{2}}}, \theta = (a,b,c) = (2, 1.5, 0.1)$$
	Lower graph shows trajectory that has been built by Euler's method and upper shows increments of stochastic process. 
	
	\begin{figure}[!htb]
		\caption{Estimation for one-parameter model.}
		\includegraphics[width=\textwidth]{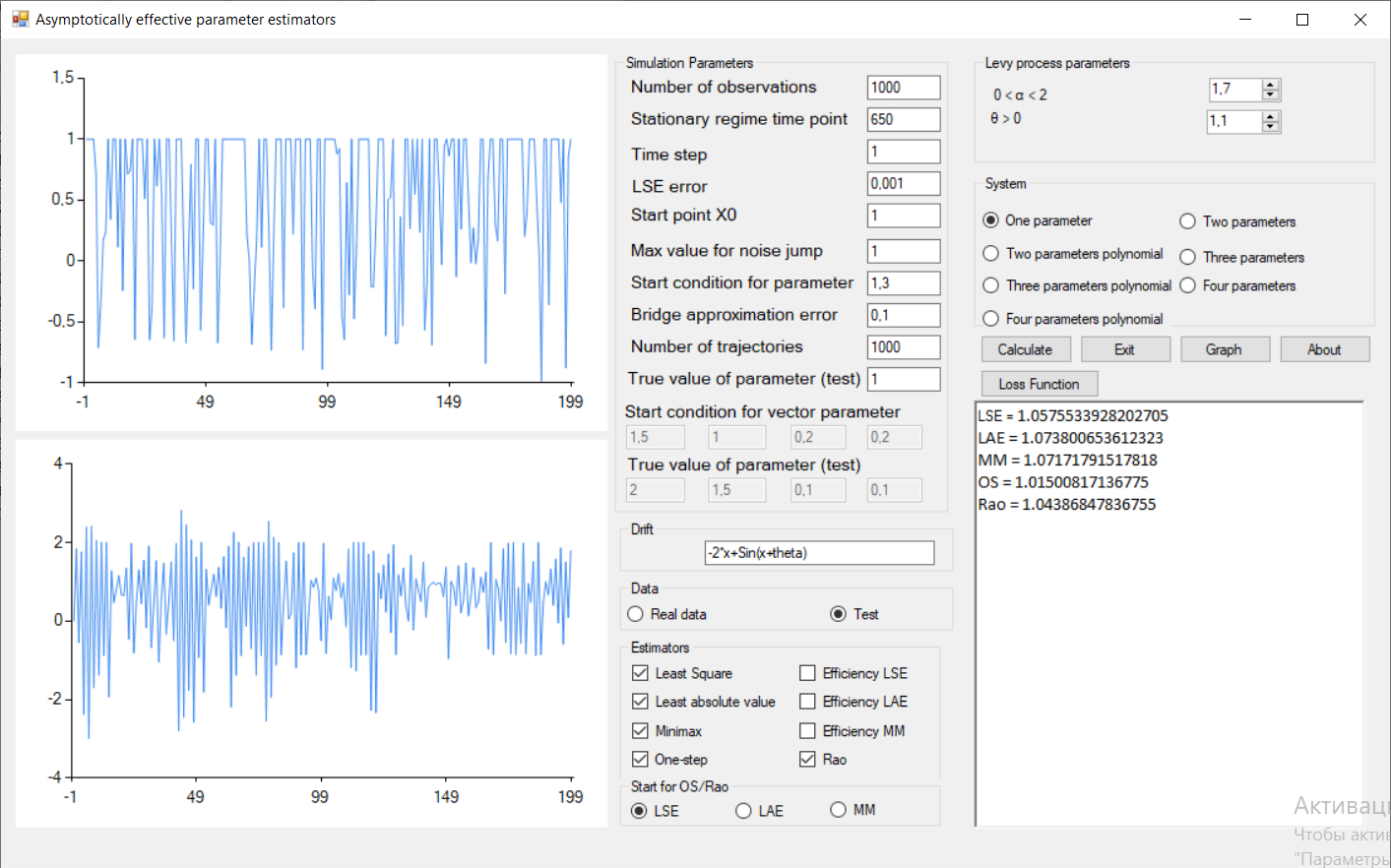}
		\label{fig4}
	\end{figure}
	
	\subsection*{One-parameter estimation}
	Consider equation (1) with  $A = -2x + \sin(x+\theta)$, true value of $\theta$ is 1, $\alpha = 1.75, T = 1000, h = 1, x_{0} = 1$, max value for noise jump = 1. 1000 experiments have been made and statistics such as mean absolute value, relative mean square error and efficiency have been calculated. 
	
	The results are given in table \ref{tab4}. For comparison, a different start for one-step and Rao estimators has been given by different $L_p$ estimators.

	\begin{table}[!htb]	
		
		\begin{center}
			
			\begin{tabulary}{\textwidth}{|c|c|c|c|c|}
				\hline 
				Estimator & Mean $\theta$ & MAD $\theta$ & RMSE $\theta$ & Efficiency\\ 
				\hline 
				$L^{1}$ & 1.0720 & 0.0721 & 0.0727 & 0.0899\\ 
				\hline 
				$L^{2}$ & 1.0608 & 0.0608  & 0.0609 & 0.1073\\ 
				\hline 
				$L^{\infty}$ & 1.1134 & 0.1406  & 0.1611 & 0.0406\\
				\hline 
				OS-$L^{1}$ & 1.0151 & 0.0152  & 0.0153 & 0.4275\\ 
				\hline 
				OS-$L^{2}$ & 1.0269 & 0.0269 &0.0278 & 0.2347\\ 
				\hline 
				OS-$L^{\infty}$ & 1.0153 & 0.0153  & 0.01687 & 0.3873\\
				\hline 
				Rao-$L^{1}$ & 1.0584 & 0.0588  & 0.0592 & 0.1105\\ 
				\hline 
				Rao-$L^{2}$ & 1.0471 & 0.0743 & 0.0632 & 0.1384\\ 
				\hline 
				Rao-$L^{\infty}$ & 1.0997 & 0.1315  &0.1517 & 0.0431\\
				\hline
			\end{tabulary} 
			\caption{Efficiency value for one-parametric model}\label{tab4}
			
		\end{center}
	\end{table}

	The efficiency of method depends on number of trajectories and number of observations but insignificantly. Efficiency grows by half of percent  with an increase in trajectories by one hundred and decreases by half of percent  with an increase in number of observation by one hundred. This effect can be explained by the fact that the step is fixed and with an increase in the number of observations, we accordingly increase the observation interval.

	\subsection*{Multiparameter estimation}
	
Consider a model given by drift function $A = -2x + \arctan(x^{2} + a) + \dfrac{b}{\sqrt{1+x^{2}}}$, $(a,b) = (1,1), \alpha = 1.75, T = 2000, h = 1, x_{0} = 1.$ 1000 experiments have been made. Table \ref{tab5} shows statistics for estimators. Here RMSE and MAD are calculated as maximum absolute value between estimated and true value.

\begin{table}[!htb]	
	
	\begin{center}
		
		\begin{tabulary}{\textwidth}{|c|c|c|c|c|}
			\hline 
			Estimator & Mean a & Mean b & MAD  & RMSE  \\ 
			\hline 
			$L^{1}$ & 1,0003 & 0,9996 & 0,0039 &  0,007\\ 
			\hline 
			$L^{2}$ & 1,0096 & 0,992  & 0,0779 & 0,1019\\ 
			\hline 
			$L^{\infty}$ & 1,0164 & 1,0107  & 0,2799 & 0,2904\\
			\hline 
			
		\end{tabulary} 
		\caption{Statistics for two-parametric model.}\label{tab5}
		
	\end{center}
\end{table}
	
	 Consider a model given by polynomial drift function $ax^{3}+bx^{2}+cx$, $(a,b,c) = (-0.2,-0.3, -0.6,)$, $\alpha = 1.75, T = 1000, h = 1, x_{0} = 1$, max value for noise jump is 1.  Table \ref{tab6} shows statistics for estimators. Here RMSE and MAD are calculated as maximum absolute value between estimated and true value.

	\begin{table}[!htb]	
		
		\begin{center}
			
			\begin{tabulary}{\textwidth}{|c|c|c|c|c|c|}
				\hline 
				Estimator & Mean a & Mean b & Mean c &  MAD  & RMSE  \\ 
				\hline 
				$L^{1}$ & -0,2246 &  -0,27703 & -0,5967 &  0,0606 &  0,0889\\ 
				\hline 
				$L^{2}$ & -0,2015 & -0,2987 & -0,6018 &  0,0666 & 0,1036\\ 
				\hline 
				$L^{\infty}$ & -0.1901 & -0.2916 & -0,6003 &  0.0241 & 0.0436\\
				\hline 
				
			\end{tabulary} 
			\caption{Statistics for model with polynomial drift.}\label{tab6}
			
		\end{center}
	\end{table}
	
	In general bias does not significantly depends on number of observations. Increasing this number by 100 gives the difference in fifth digit. 
	
	\section{Conclusion}
	It has been determined that for parameter estimation in models with alpha-stable noise numerical hybrid method give higher precision. A combination of such methods, which has not been used before, shows good results as shown in Table 1. The experiments have shown that one step and Rao estimators improve the accuracy of estimators. The efficiency algorithm is implemented and programmed and with the help of it a comparative analysis of such estimators as $L^{p}$, one-step and Rao. In practice, quite expected effects have been confirmed. One step and Rao estimators improve value, the growth of trajectories leads to an increase in efficiency, an increase in observations with a fixed step leads to the efficiency decreasing.

	\end{document}